\documentclass[%
 reprint, 
 amsmath,amssymb,
 aps, 
 prl,
floatfix,
]{revtex4-2}

\usepackage{graphicx}
\usepackage{dcolumn}
\usepackage{bm}
\usepackage{upgreek}
\usepackage{textcomp}

\usepackage[colorlinks=true,linkcolor=blue,anchorcolor=blue,citecolor=blue,urlcolor=blue]{hyperref}

\begin{document}

\preprint{APS/123-QED}

\title{Clock precision beyond the Standard Quantum Limit at \texorpdfstring{$10^{-18}$}{10⁻¹⁸} level
}%


\author{Y. A. Yang}
\thanks{These authors contributed equally to this work.}
\author{Maya Miklos}
\thanks{These authors contributed equally to this work.}
\author{Yee Ming Tso}
\author{Stella Kraus}
\author{Joonseok Hur}
\author{Jun Ye}%
 \email{Ye@jila.colorado.edu}
\affiliation{%
 JILA, National Institute of Standards and Technology and the University of Colorado, Boulder, Colorado 80309-0440, USA \\
and Department of Physics, University of Colorado, Boulder, Colorado 80309-0390, USA
}%



\date{\today}

\begin{abstract}
Optical atomic clocks with unrivaled precision and accuracy have advanced the frontier of precision measurement science and opened new avenues for exploring fundamental physics. A fundamental limitation on clock precision is the Standard Quantum Limit (SQL), which stems from the uncorrelated projection noise of each atom. State-of-the-art optical lattice clocks interrogate large ensembles to minimize the SQL, but density-dependent frequency shifts pose challenges to scaling the atom number. The SQL can be surpassed, however, by leveraging entanglement, though it remains an open problem to achieve quantum advantage from spin squeezing at state-of-the-art stability levels. Here we demonstrate clock performance beyond the SQL, achieving a fractional frequency precision of 1.1 $\times 10^{-18}$ for a single spin-squeezed clock. With cavity-based quantum nondemolition (QND) measurements, we prepare two spin-squeezed ensembles of $\sim$30,000 strontium atoms confined in a two-dimensional optical lattice. A synchronous clock comparison with an interrogation time of 61 ms achieves a metrological improvement of 2.0(2) dB beyond the SQL, after correcting for state preparation and measurement errors. These results establish the most precise entanglement-enhanced clock to date and offer a powerful platform for exploring the interplay of gravity and quantum entanglement.

\end{abstract}

\maketitle

State-of-the-art optical lattice clocks (OLC) now deliver frequency precision and accuracy below $10^{-18}$~\cite{Ludlow2015Rev.Mod.Phys.a, Bloom2014Nature, Campbell2017Science, Marti2018Phys.Rev.Lett.a, Beloy2021Nature, Zheng2022Nature,bothwellNature2022,Aeppli2024Phys.Rev.Lett., Li2024Metrologia, Hausser2025Phys.Rev.Lett.}, enabling tests of fundamental physics~\cite{Chou2010Science, Safronova2018Rev.Mod.Phys.,langePhys.Rev.Lett.2021, bothwellNature2022}, providing promising candidates for redefining the SI second~\cite{Riehle2018Metrologia,Beloy2021Nature,Dimarcq2024Metrologia}, and inspiring applications across diverse fields~\cite{Takano2016NaturePhoton, McGrew2018Nature,Ye2024Phys.Rev.Lett.}. Recent synchronous comparisons~\cite{Campbell2017Science,Marti2018Phys.Rev.Lett.a,Young2020Nature,Zheng2022Nature,bothwellNature2022,Zheng2024Phys.Rev.X} of two clocks reject local oscillator (Dick) noise, pushing measurement precision into $10^{-21}$ decade~\cite{bothwellNature2022}. Yet clock instability is fundamentally limited by the SQL, which scales as $\sigma_{\text{SQL}} \sim 1/\sqrt{N}$ due to the projection noise of the $N$ uncorrelated atoms. Classical OLCs chase this asymptotic scaling by maximizing $N$, and battle density-dependent frequency shifts by engineering the lattice geometry to cancel~\cite{sr1density}, or separately resolve~\cite{swallows, Campbell2017Science} interaction effects. Introducing entanglement offers a route beyond this bound: Redistributing spin-noise uncertainty improves the measurement precision for fixed $N$~\cite{winelandPhys.Rev.A1992, kitagawaPhys.Rev.A1993, winelandPhys.Rev.A1994, schleier-smithPhys.Rev.Lett.2010, Cox2016Phys.Rev.Lett., Hosten2016Nature,Pezze2018Rev.Mod.Phys.,Pedrozo-Penafiel2020Nature,Marciniak2022Nature,Eckner2023Nature,Robinson2024Nat.Phys.,Cao2024Nature}. Over the past decade, entangled resource states have been generated on both microwave and optical transitions, across platforms ranging from cold atom ensembles~\cite{schleier-smithPhys.Rev.Lett.2010,Cox2016Phys.Rev.Lett.,Hosten2016Nature,Pedrozo-Penafiel2020Nature,Robinson2024Nat.Phys.}, trapped ions~\cite{Marciniak2022Nature}, and tweezer-controlled atom arrays~\cite{Eckner2023Nature,Cao2024Nature}. Nevertheless, achieving sub-SQL precision measurements at the $10^{-18}$ frontier remains challenging, as the fragility of entanglement demands ever more sophisticated quantum control and engineering.

In this Letter, we report a clock comparison at the $10^{-18}$ precision level, also demonstrating a metrological gain of 2.0(2) dB beyond the SQL after correcting for state preparation and
measurement (SPAM) errors. We prepare two spin-squeezed ensembles of $^{87}$Sr atoms in a two-dimensional (2D) OLC which is integrated with collective strong-coupling cavity quantum electrodynamics (QED) for QND measurements~\cite{Robinson2024Nat.Phys.}. Enhanced control of atomic motion in the 2D lattice enables significant reduction in decoherence induced by QND measurements, and more precise control of the clock laser increases the spin rotation fidelity. The synchronous clock comparison achieves a fractional frequency instability of $8.0(2)\times10^{-17}/\sqrt{\tau}$ (where $\tau$ is the averaging time in seconds), resulting in precision of $1.6\times10^{-18}$ for the full measurement time.

\begin{figure}[htb!]
    \includegraphics[width=8.6cm]{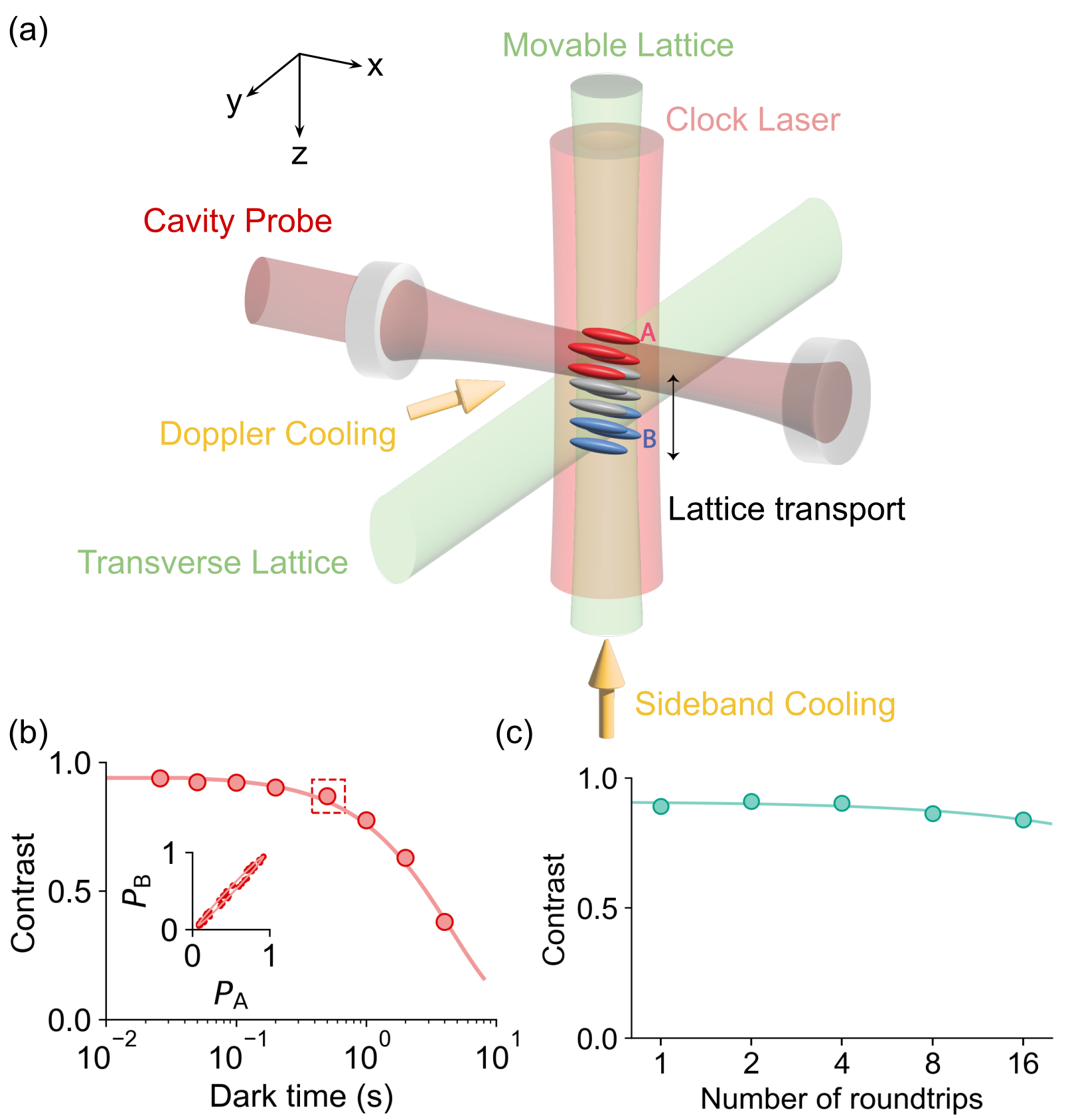}
    \caption{Enhanced control of atomic motion. 
    (a) Overview of the setup. Atoms are trapped in a 2D optical lattice formed by a movable lattice along the vertical ($z$) direction and a transverse lattice along $y$. Relative detuning of the movable lattice beams enables transport of atoms along $z$ direction, allowing independent cavity-based QND measurements of subensembles A (red tubes) and B (blue tubes). The transverse lattice enhances confinement and atom-cavity coupling homogeneity. A clock laser propagating along $z$ globally drives the clock transition. Combined with sideband cooling along $z$ and Doppler cooling in the $x-y$ plane, atomic temperatures below 0.5 $\upmu$K are achieved. 
(b) Two-ensemble Ramsey contrast as a function of dark time, demonstrating an atom-atom coherence time of $4.5(1)$~s. The inset shows the parametric plot of the excitation fraction $P_{\text{A}}$ and $P_{\text{B}}$ for subensembles A and B at 500 ms dark time.
(c) Contrast decay as a function of the number of lattice transport roundtrips, measured at a dark time of 141 ms. 
}
    \label{fig:1}
\end{figure}

Our system incorporates a state-of-the-art OLC with a collective cavity QED system, as first demonstrated in Ref.~\cite{Robinson2024Nat.Phys.}. $^{87}$Sr atoms are prepared in a two-stage magneto-optical trap~\cite{katoriPhys.Rev.Lett.1999} located at $\sim40$~mm below the cavity, where atoms are subsequently loaded into a movable lattice formed by two counter-propagating beams intersecting the cavity mode. By introducing a detuning between two lattice beams, the movable lattice transports atoms along the vertical direction (Fig.~\ref{fig:1}(a)), allowing arbitrary sections of the atomic ensemble to be positioned within the cavity (mode waist $\sim71 \ \upmu$m)~\cite{Sauer2004Phys.Rev.A}. A transverse optical lattice, which is elongated along the vertical direction, ensures a uniform confinement across the whole atomic ensemble. These two lattices provide two-dimensional confinement in the cavity region while still allowing the longitudinal atomic motion along the cavity axis. After applying sideband and Doppler cooling, the average vibrational occupations along two lattice directions are below 0.1. 

The clock laser~\cite{mateiPhys.Rev.Lett.2017} propagating along the movable lattice globally drives the whole atomic ensemble on the $\vert{}^1S_0,\ m_F = -9/2\rangle\rightarrow \vert{}^3P_0,\ m_F = -9/2\rangle$ transition. We select two independent subensembles of atoms, labeled A and B, that are separated by 140 $\upmu$m for clock comparison. The enhanced control of atomic motion through the 2D lattice and cooling results in an atom-atom coherence time of $4.5(1)$~s (Fig.~\ref{fig:1}(b)). Furthermore, lattice transports during Ramsey interrogation disturb the coherence only slightly. We characterize this by measuring the Ramsey contrast loss after multiple sets of upward and downward transports. Each transport covers a distance of 140 $\upmu$m in 2.5 ms. As shown in Fig.~\ref{fig:1}(c), at 141 ms interrogation time, the contrast decreases from 91(1)\% to 84(2)\% after 16 roundtrips. Combined with cavity-based QND measurements, this capability enables both mid-circuit transports and nondestructive mid-circuit measurements~\cite{Beugnon2007NaturePhys,Gehr2010Phys.Rev.Lett.,Bluvstein2022Nature,Norcia2023Phys.Rev.X} during clock operation, which are essential for the spin-squeezed clock comparison demonstrated here. For the spin-squeezed clock comparison, we implement two roundtrips during the Ramsey interrogation and additional two roundtrips afterward.



\begin{figure}[t]
    \includegraphics[width=8.6cm]{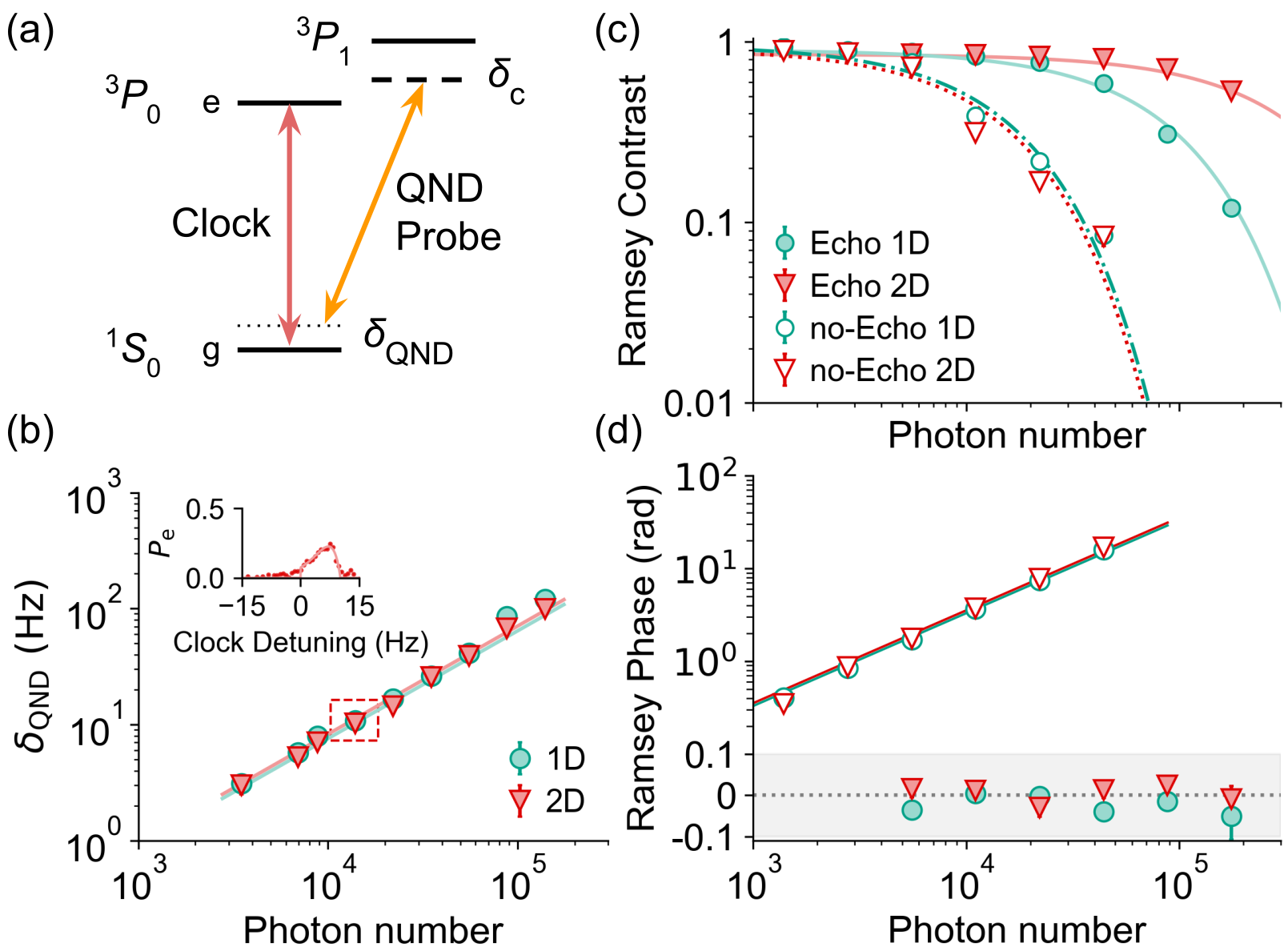}
    \caption{Characterization of QND measurements. 
    (a) The cavity resonance is detuned from the ${}^1S_0 \rightarrow {}^3P_1$ transition 
    by $\delta_{\text{c}}/(2\pi)=-4\ \text{MHz}$. The probe laser for cavity-based QND measurements induces inhomogeneous light shifts $\delta_{\text{QND}}$ on the ${}^1S_0 \rightarrow {}^3P_0$ clock transition due to non-uniform atom-cavity coupling.
    (b) Measured QND-induced light shifts $\delta_{\text{QND}}$ using Rabi spectroscopy in two lattice configurations: 1D (movable lattice only, 19 $E_{\text{r}}$ depth) and 2D (movable lattice at 19 $E_{\text{r}}$ plus transverse lattice at 12 $E_{\text{r}}$). The inset shows a broadened, distorted line shape resulting from light shifts.
    (c,d) Inhomogeneous light shifts lead to loss of coherence and phase shift in Ramsey spectroscopy, which can be greatly suppressed with spin echo. Ramsey contrast with and without echo sequence is compared in two lattice configurations, showing that enhanced control of atomic motion in the 2D lattice improves spin-echo performance. The phase shift induced by QND probing with spin echo is negligible.
}
    \label{fig:2}
\end{figure}

The collective atom-cavity coupling enables QND measurements of ground-state atom number $N_{\text{g}}$ in the cavity~\cite{Pezze2018Rev.Mod.Phys.}. We detune the cavity from the $\vert{}^1S_0,\ m_F = -9/2\rangle\rightarrow {}\vert^3P_1,\ m_F = -11/2\rangle$ transition by $\delta_{\text{c}}/(2\pi)=-4$~MHz (Fig.~\ref{fig:2}(a)). An ensemble of $N_{\text{g}}$ atoms induces a dispersive shift of cavity resonance by $\delta\omega_{\text{c}}(N_\text{g}) =N_\text{g}\frac{g^2}{\delta_{\text{c}}}$~\cite{Berman1994}, which is measured by balanced optical homodyne detection. Here, $2g = 2\pi \times (2\times5.1(2))$ kHz is the effective vacuum Rabi frequency. A clock $\pi$ pulse exchanges the ground- and clock-state populations, followed by a second QND measurement that determines the number of atoms initially in the clock state, $N_\text{e}$. Thus, two QND measurements with a clock $\pi$ pulse provide a nondestructive measurement of collective spin projection $J_z = (N_\text{g} - N_\text{e})/2$ on the clock transition. The quantum projection noise (QPN) for a coherent spin state (CSS) on the equator of the Bloch sphere is $\Delta J_{z}^{\text{QPN}} = \sqrt{N}/2$, where $N=N_\text{g} + N_\text{e}$ and $\Delta$ denotes the standard deviation~\cite{Itano1993Phys.Rev.A}.

The same atom-cavity interaction, however, induces loss of coherence via light shifts~\cite{lerouxPhys.Rev.Lett.2010,Cox2016Phys.Rev.Lett.,Hosten2016Nature,Bowden2020Phys.Rev.X}. The QND probe laser, on resonance with the cavity, causes light shifts on the clock transition. Although the $J_z$ measurement naturally implements a spin-echo~\cite{Hahn1950Phys.Rev.} sequence to cancel out the light shift, the atomic motions and finite temperature degrade the spin-echo cancellation. To systematically study this effect, we measure the QND-induced light shift $\delta_{\text{QND}}$ on the clock transition with Rabi spectroscopy. The photon number cited in this work refers to that incident on cavity in a 40-ms QND probe pulse. The distorted line shape (Fig.~\ref{fig:2}(b)) reveals spatial inhomogeneity in atom-cavity interaction. Consequently, a single QND measurement applied during the Ramsey interrogation imposes an inhomogeneous phase shift and reduces the coherence of the ground–clock superposition, as illustrated by ``no-Echo" traces in Fig.~\ref{fig:2}(c, d). By applying a spin-echo sequence ($J_z$ measurement described above), Ramsey coherence can be restored with negligible phase shift. We further compare the effect of spin echo in two lattice configurations: (i) the conventional 1D lattice used in state-of-the-art OLCs and (ii) a 2D lattice~\cite{swallows} that provides tighter confinements transverse to the cavity axis. As shown in Fig.~\ref{fig:2}(c), the 2D lattice preserves the Ramsey coherence more effectively, as the longitudinal atomic motion along the cavity axis averages over the standing-wave cavity mode efficiently. In contrast, radial motion in the 1D lattice averages over not only the standing-wave but also the Gaussian envelope of the cavity mode, leading to reduced coherence.

\begin{figure}[t]
    \includegraphics[width=8.6cm]{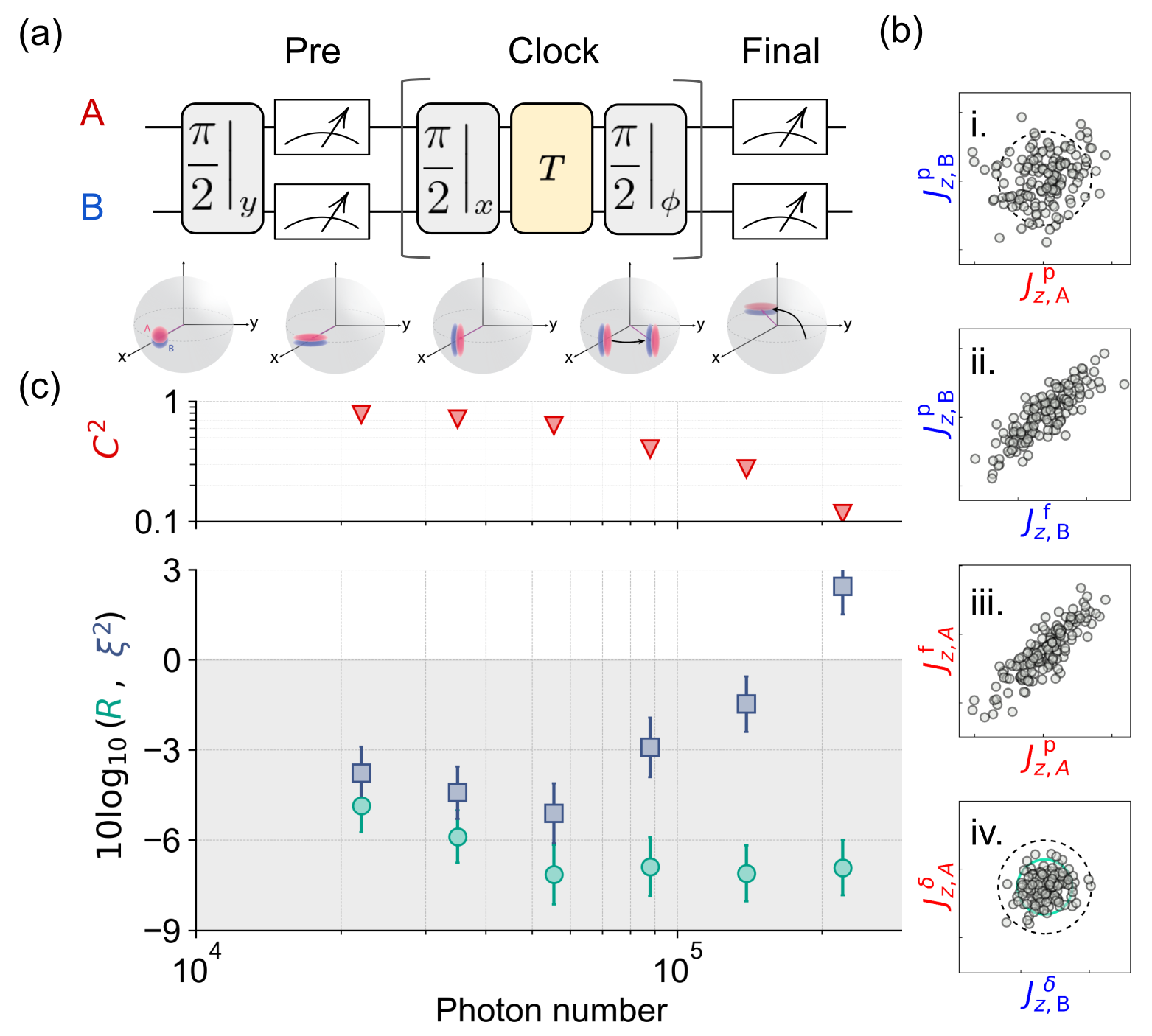}
    \caption{Spin-noise reduction and metrological enhancement.
    (a) Measurement sequence. The ``Pre" $J_z$ measurements project the CSSs into conditionally spin-squeezed states. The clock interrogation sequence consists of two $\pi/2$ pulses separated by an interrogation time $T$. ``Final" $J_z$ measurements readout the spin-squeezed states after interrogation.
    (b) Correlations among successive $J_z$ measurements. Dashed circles with radius $2 \Delta J_{z, \text{A(B)}}^{\text{QPN}}$ indicate the QPN-limited scatter, independently calibrated from atom-cavity coupling. 
    (c) Metrological enhancement as a function of probe photon number. Upper panel: the squared effective contrast $C^2=C_{\text{f}}^2/C_{\text{i}}$ versus photon number. Lower panel: spin noise reduction $R$ and metrological gain $\xi^2 = R/C^2$. A maximum spin-noise reduction of $R = -7.2(1.0)$ dB is observed. At the optimal photon number, a metrological enhancement $\xi^2=-5.1(1.0)$ dB is achieved.
}
    \label{fig:3}
\end{figure}

Armed with improved Ramsey coherence, we benchmark the metrological usefulness of the spin-squeezed states (SSSs) generated by QND measurements. Following the experimental sequence shown in Fig.~\ref{fig:3}(a), ensembles A and B are first initialized in the clock excited state; then a clock $\pi/2$ pulse prepares a CSS on the equator of the Bloch sphere, establishing the clock coherence. Subsequent $J_z$ measurements (``Pre" or ``p") project ensembles A and B into conditionally spin-squeezed states~\cite{Cox2016Phys.Rev.Lett.}, respectively. Repeated $J_z$ measurements (``Final" or ``f") show correlations with ``Pre" measurements below the QPN level for both ensembles, as shown in Fig.~\ref{fig:3}(b). This allows the quantum noise to be partially canceled in the difference $J_{z,\text{A(B)}}^{\delta } = J_{z,\text{A(B)}}^{\text{f}}-\beta_{\text{A(B)}}J_{z,\text{A(B)}}^{\text{p}}$, where optimal estimator $\beta_{\text{A(B)}} < 1$ accounts for uncorrelated technical noise. 
We define the two-ensemble spin-noise reduction as 

\begin{equation}
\label{eq:R}
     R = \left(\frac{\Delta \left(J_{z, \text{A}}^{\delta} -J_{z, \text{B}}^{\delta}\right)}{\sqrt{N_\text{A}+N_\text{B}}/2}\right)^2,
\end{equation}
where $\sqrt{N_\text{A}+N_\text{B}}/2$ is the combined QPN for $J_{z,\text{A(B)}}$. 

The metrological usefulness of spin-noise reduction has to compensate for the contrast loss from ``Pre" $J_z$ measurements during Ramsey interrogation. The spin-squeezed states generated by $J_z$ measurements are rotated into phase-sensitive orientation during clock interrogation and a final $\pi/2$ pulse maps the phase onto $J_z$~\cite{Robinson2024Nat.Phys.}. We quantify this trade-off using effective contrast, defined as $C = C_{\text{f}}/\sqrt{C_{\text{i}}}$, where $C_{\text{i}}$ ($C_{\text{f}}$) is the two-ensemble average Ramsey contrast without (with) ``Pre" $J_z$ measurements. The initial contrast, $C_{\text{i}} = 82(1)\%$ is limited by the SPAM errors accumulated over multiple clock rotations, while the contrast loss in $C_{\text{f}}$ is primarily from the imperfect spin-echo cancellation and free space scattering. The metrological gain of spin-squeezed states is characterized by the squeezing parameter $\xi^2 = R/C^2$, where $\xi^2 < 1$ serves as an entanglement witness ~\cite{lerouxPhys.Rev.Lett.2010a, Chen2011Phys.Rev.Lett., Cox2016Phys.Rev.Lett.} and $\xi^2 = 1$ defines the SQL after correcting for SPAM errors. For reference, the Wineland parameter $\xi_{\text{W}}^2 = \xi^2/C_{\text{i}}$~\cite{winelandPhys.Rev.A1992} quantifies metrological enhancement without accounting for initial contrast; in our case $C_{\text{i}}$ corresponds to a 0.9(1) dB correction. At the optimal number of photons for QND measurements, we achieve a two-ensemble spin-noise reduction of -7.2(1.0) dB and metrological gain of -5.1(1.0) dB.

\begin{figure}[t]
    \includegraphics[width=8.6cm]{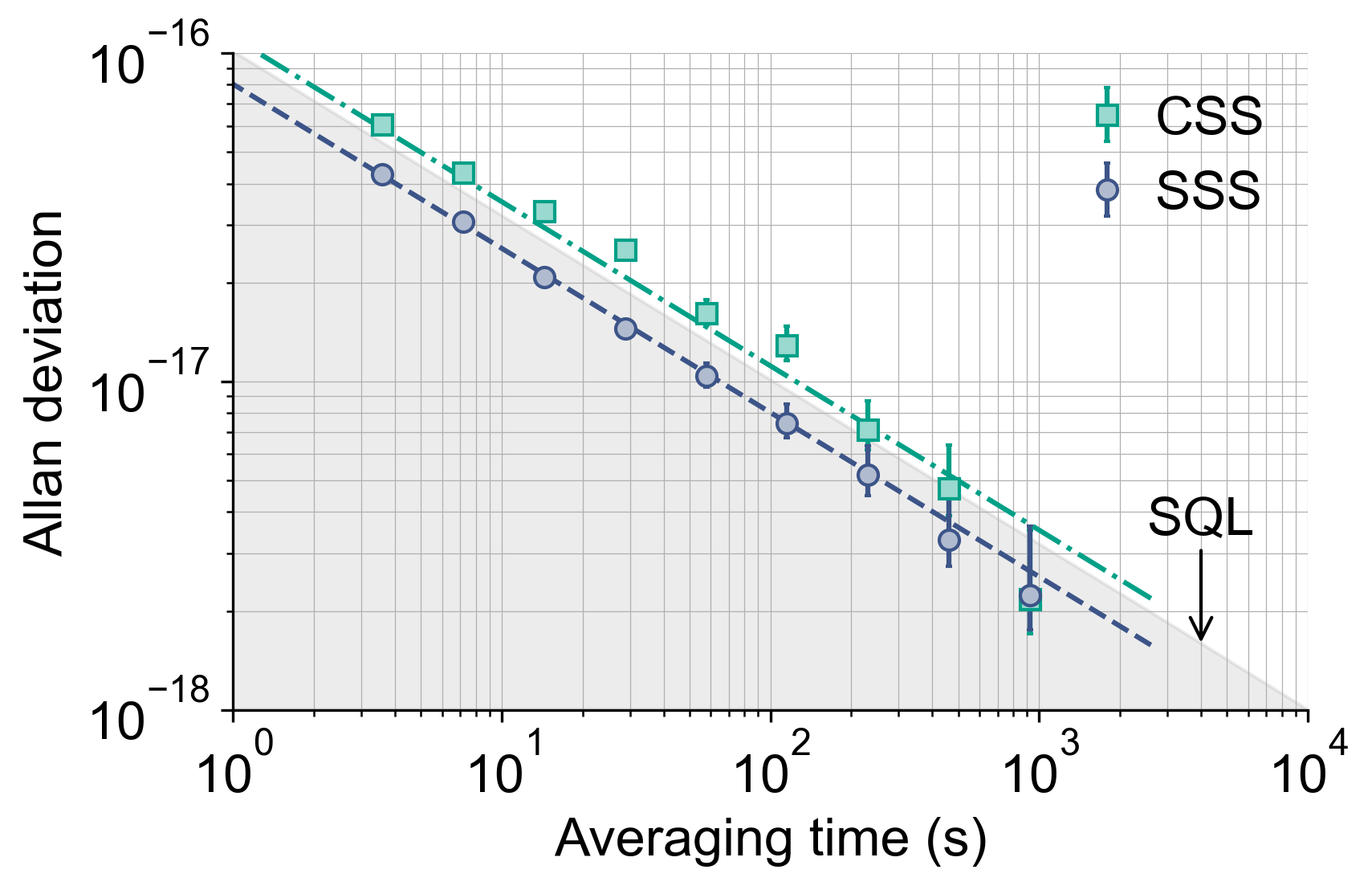}
    \caption{Synchronous clock comparison.
    With the movable lattice, ensembles A and B are alternately transported into the cavity for QND measurements, enabling direct synchronous clock comparison. Since the clock laser addresses two ensembles globally, common-mode clock laser noise is largely rejected. The measured frequency instability of CSS-CSS comparison (green points) agree with the theoretical QPN limit. In the meantime, the SSS-SSS comparison (blue points) not only achieves a 3.3(2) dB enhancement relative to the CSS-CSS case, but also demonstrates a 2.0(2) dB metrological advantage beyond the Standard Quantum Limit (SQL), after correcting for SPAM errors (initial contrast $C_{\text{i}} = 82(1)\%$ corresponds to a 0.9(1) dB correction to SQL). The synchronous clock comparison achieves a fractional frequency instability of $8.0(2)\times10^{-17}/\sqrt{\tau}$, resulting in a comparison precision of $1.6\times10^{-18}$ for full measurement time. Allan deviations of the frequency difference between two ensembles are plotted after subtracting a linear frequency drift of 1.6 $\upmu$Hz/s. Error bars represent $1\sigma$ statistical confidence interval.}

    \label{fig:4}
\end{figure}

Furthermore, to directly demonstrate the metrological gain in clock performance, we perform clock comparisons between two ensembles~\cite{Campbell2017Science,Marti2018Phys.Rev.Lett.a,Young2020Nature,Zheng2022Nature,bothwellNature2022,Zheng2024Phys.Rev.X}. The clock interrogation time $T$ is 61 ms. As shown in Fig.~\ref{fig:4}, the clock comparison with CSSs (without ``Pre" $J_z$ measurements) shows a fractional frequency instability of $1.18(2)\times10^{-16}/\sqrt{\tau}$, consistent with the QPN limit~\cite{Robinson2024Nat.Phys.} set by atom number and non-unitary $C_{\text{i}}$. The clock comparison with SSSs achieves a 3.3(2) dB reduced instability of $8.0(2)\times10^{-17}/\sqrt{\tau}$, corresponding to a 2.0(2) dB metrological gain beyond the SQL, after correcting for SPAM errors. After a full measurement time of 43 min, we achieve a single-clock fractional frequency uncertainty of $1.1\times10^{-18}$, establishing the most precise entanglement-enhanced clock to date. We note that a linear frequency drift of 1.6 $\upmu$Hz/s, attributed to a long-term magnetic field gradient drift, is subtracted when plotting Allan deviation of the frequency difference between two ensembles.

The observed metrological gain in clock performance is limited by noise introduced during optical rotations. In addition to the non-unitary $C_{\text{i}}$, the two $\pi/2$ pulses to rotate the squeezed-state orientation and map the clock phase into populations can introduce excess noise from the anti-squeezing quadrature~\cite{Hosten2016Nature,Cox2016Phys.Rev.Lett.}. Achieving high-fidelity optical rotations requires a low-noise optical local oscillator~\cite{mateiPhys.Rev.Lett.2017,Oelker2019Nat.Photonics} and high-fidelity quantum state engineering~\cite{Ye2008Science,Katori2011NaturePhoton,Campbell2017Science,Lis2023Phys.Rev.X,Madjarov2020Nat.Phys.}. Further improvements can be realized by confining atoms in a three-dimensional optical lattice~\cite{Campbell2017Science}, where tight spatial localization enables better control of atomic motions and interactions. In addition, quantum sensing protocols, such as signal amplification via time-reversed interactions~\cite{Davis2016Phys.Rev.Lett.,Linnemann2016Phys.Rev.Lett.,Frowis2016Phys.Rev.Lett.,Nolan2017Phys.Rev.Lett.,Colombo2022Nat.Phys.} or quantum variational optimization~\cite{Kaubruegger2019Phys.Rev.Lett.,Kaubruegger2021Phys.Rev.X, Marciniak2022Nature}, offer promising paths for enhanced metrological performance. 

In conclusion, we have demonstrated clock performance beyond the SQL, achieving a milestone precision at the $10^{-18}$ level. These results can be further improved through more advanced quantum control and state engineering. This work lays the foundation for future evaluations of systematic effects in entanglement-enhanced clocks at $10^{-18}$ and below, and provides a powerful platform for exploring the interplay of gravity and quantum entanglement~\cite{bothwellNature2022,Chu2025Phys.Rev.Lett.}, as well as quantum many-body physics in entangled systems~\cite{Amico2008Rev.Mod.Phys.,Martin2013Science, Qu2019Phys.Rev.A,Hutson2024Science}.

\textit{Acknowledgments--} We thank B. Heizenreder and K. Kim for technical contributions, and J. K. Thompson, W. J. Eckner, and D. Young for useful discussions and review of this manuscript. Funding is provided by the US Department of Energy, Office of
Science, National Quantum Information Science Research Centers, Quantum Systems Accelerator. Additional funding is provided by the National Science Foundation QLCI OMA-2016244, V. Bush Fellowship, JILA Physics Frontier Center PHY-2317149, and the National Institute of Standards and Technology. M.M. is supported by an NSF Graduate Research Fellowship. 

\bibliography{Sr3}

\end{document}